\documentclass[prb,twocolumn,showpacs,amsmath,amssymb]{revtex4}
\usepackage{graphicx}
\usepackage{dcolumn}
\usepackage{bm}
\usepackage{epsfig}
\begin{document}


\title{The kagome staircase compounds Ni$_{3}$V$_{2}$O$_{8}$ 
and Co$_{3}$V$_{2}$O$_{8}$ studied with implanted muons}

\author{T. Lancaster}
\email{t.lancaster1@physics.ox.ac.uk}
\author{S. J. Blundell}
\author{P. J. Baker}
\author{D. Prabhakaran}
\author{W. Hayes}
\affiliation{
Clarendon Laboratory, Oxford University Department of Physics, Parks
Road, Oxford, OX1 3PU, UK
}
\author{F. L. Pratt}
\affiliation{
ISIS Facility, Rutherford Appleton Laboratory, Chilton, Oxfordshire OX11 0QX, UK}

\date{\today}

\begin{abstract}
We present the results of muon-spin relaxation ($\mu^{+}$SR)
measurements on the kagome staircase compounds
Ni$_{3}$V$_{2}$O$_{8}$ and Co$_{3}$V$_{2}$O$_{8}$. 
The magnetic behavior of these materials may be described in
terms of two inequivalent magnetic ion sites, known as spine sites
and cross-tie sites. Our $\mu^{+}$SR results allow us to
probe each of these sites individually, revealing the distribution
of the local magnetic fields near these positions.
We are able not only to confirm the magnetic structures of the various
phases proposed on the basis
of bulk measurements but also to give an insight into the temperature
evolution of the local field distribution in each phase. 
\end{abstract}

\maketitle

\section{Introduction}

Geometric frustration often gives rise to highly degenerate ground
state manifolds with unusual spin correlations. Small perturbations 
become
decisive in lifting the degeneracy and may lead to the existence 
of rich phase diagrams\cite{moessner}. One of
the most studied frustrated systems is the two-dimensional (2D) 
kagome lattice.  This is formed from corner sharing triangles
of spins with equal antiferromagnetic (AFM) interactions between
nearest neighbours. The expected low temperature spin state 
ranges from a quantum spin liquid for the $S=1/2$ Heisenberg
model\cite{sachdev} 
to  $\sqrt{3} \times \sqrt{3}$ long range magnetic order (LRO)
for the classical ($S \rightarrow \infty$) case\cite{huse}. Real systems 
approximating the kagome lattice 
show a variety of ground states
resulting from the presence of
relatively weak interactions.
Notable examples 
include $S=3/2$
SrCr$_{9}$Ga$_{3}$O$_{19}$ which displays a spin liquid phase and
short range $\sqrt{3} \times \sqrt{3}$
order\cite{ramirez,lee}, and the jarosite materials whose interlayer 
Dzyaloshinskii-Moriya
interactions give rise to a number of commensurate magnetic 
structures\cite{wills}.

\begin{figure}
\begin{center}
\epsfig{file=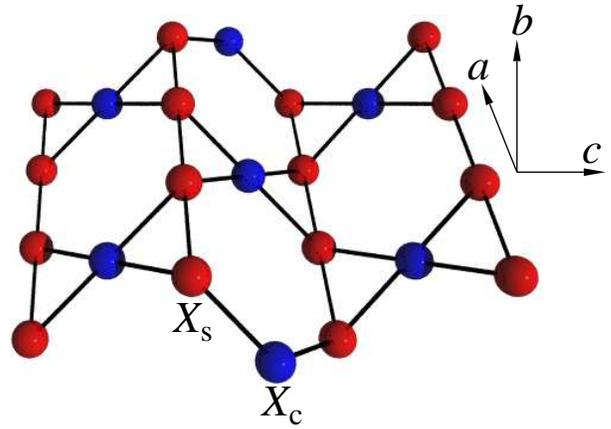,width=8cm}
\caption{(Color online) Crystal structure of the $X$VO kagome staircase layers
showing the sites of $X^{2+}$ spine spins ($X_{\mathrm{s}}$) in red
and the cross-tie spins ($X_{\mathrm{c}}$) in blue.
\label{pov}}
\end{center}
\end{figure}

Another class of material that has been the subject of much recent
research effort is
 $X_{3}$V$_{2}$O$_{8}$ ($X$VO), where $X$
 includes\cite{rogado,rogado2} 
Ni ($S=1$), Co ($S=3/2$) and Cu ($S=1/2$). This system is a further
variant of the 2D
kagome paradigm, which has attracted attention both for its complex
magnetic
phases\cite{rogado,lawes_mag,kenzelmann,chen,szymczak,wilson1,wilson2} 
and, in the case of $X$=Ni, for the appearance of ferroelectricity that
accompanies one of its magnetic phase
transitions\cite{lawes_el,harris},
making it a further example of a multiferroic material
(for a review of multiferroic behavior, see 
Ref.\onlinecite{khomskii}).
The material is formed from $X^{2+}$ spins arranged on 
buckled kagome planes (which retain the connectivity
of normal kagome planes) forming so-called kagome staircases
which are stacked along the crystallographic $b$-direction.
The staircase layers are shown
in Fig.~\ref{pov}.
The spins in the layers may be split into two 
inequivalent groups: spine spins $X_{\mathrm{s}}$ (which lie
on lines (or spines) running parallel to the $a$-direction, where the 
treads meet
the risers of the staircase) and
cross-tie spins $X_{\mathrm{c}}$ (which lie at the centre of the
treads and risers)\cite{lawes_mag}. 
The most frequently studied of the $X$VO systems have been 
Ni$_{3}$V$_{2}$O$_{8}$ (NVO) and  Co$_{3}$V$_{2}$O$_{8}$ (CVO). Despite
these materials having identical crystal symmetry and similar structural
parameters their magnetic properties are quite different. Both compounds
are frustrated and it is small perturbations that relieve the frustration
and therefore determine their behavior. As a result, the nature of the
magnetic order of the Ni$^{2+}$ spins and the Co$^{2+}$ spins is different. 

Although the details of the magnetic phases of NVO and CVO
are complex (see below), some general observations may be made on the
basis of experiment and of theoretical models\cite{chen}. 
The ordering in both materials involves spin components aligned, 
for the most part, along the $a$-direction (i.e.\ parallel
to the spines). The magnetic structure always involves ordering of
the spine spins (and may or may not also involve ordering of the
cross-tie spins). For CVO the ordering along each spine is
ferromagnetic in all phases that show LRO. For NVO, a competition
between nearest- and 
next-nearest-neighbors (due to alternating Ni-O-Ni bond
angles along the staircase) causes the ordering along the
spines to be incommensurate before adopting an antiferromagnetic
structure at low temperatures. The magnetic ordering of the system
then follows from the interaction of inter- and intralayer spines. 
The separation of the magnetic behavior of the spine and cross-tie spins
is clearly of central importance in elucidating the magnetic properties of
this system. The difference between the two materials may, in part, 
be attributed to the magnetocrystalline anisotropy, which is found to be
far larger in CVO than in NVO\cite{szymczak}.

A range of experimental techniques has been used to 
probe the bulk properties of $X$VO, most notably neutron 
diffraction\cite{lawes_mag,kenzelmann}.
However, a study of the local magnetic properties of the system has, hitherto,
been lacking. 
In this paper we present a zero field muon-spin relaxation (ZF
$\mu^{+}$SR) study of NVO and CVO. 
In recent years, implanted muons have been successful in probing
frustration related behavior in several systems from a local
viewpoint, allowing new insights into the magnetic 
properties\cite{musr1,musr2,musr3,musr4,musr5,musr6,musr7,musr8}.
Using muons we are able to probe the spine and cross-tie magnetic
sites individually, relating the observed magnetic transitions to the behavior
at each magnetic site. Our measurements allow us to confirm the
magnetic structures inferred on the basis of previous neutron studies, 
as well as to show the evolution of the internal magnetic fields
of the systems as a function of temperature. 

\section{Experimental details}

Polycrystalline samples of NVO and CVO were prepared by the
solid state reaction technique using high purity  NiO, Co$_{2}$O$_{3}$
and 
V$_{2}$O$_{5}$ chemicals.  Stoichiometric quantities were
mixed and reacted in
air at 850~$^{\circ}$C for 24~h.  The samples were reground and 
sintered (Co compound at 1050$^{\circ}$C, Ni compound at
950$^{\circ}$C) in air for 48~h. 
The phase
purity of the polycrystalline samples was checked using x-ray powder
diffraction.  The final powders were isostatically pressed into rods of
10~mm diameter and length 100~mm.  The rods were sintered in air at 
1050~$^{\circ}$C
and 950~$^{\circ}$C for the Co and Ni compounds respectively. 
 Crystal growth was achieved using an optical floating zone furnace 
(Crystal System Inc.) at a growth speed of $\sim 2$~mm/h with the seed and
feed rods counter rotating at 25~rpm in an argon/oxygen mixed gas atmosphere.
For further details see Ref.~\onlinecite{balakrishnan}.

ZF $\mu^{+}$SR  measurements were made 
on mosaics of single crystals of NVO and CVO, 
using the GPS instrument at the Swiss Muon Source, 
Paul Scherrer Institute, Villigen, Switzerland.
Six crystallites of typical volume 50~mm$^{3}$
were aligned on silver foil
such that the initial muon-spin polarization
lay along the $b$ direction. The crystal facets facing the muon
beam were of typical area 20~mm$^{2}$ and covered approximately
80\% of the area illuminated by the muon beam.
In a $\mu^{+}$SR experiment, spin-polarized
positive muons are stopped in a target sample, where the muon usually
occupies an interstitial position in the crystal.
The observed property in the experiment is the time evolution of the
muon spin polarization, the behavior of which depends on the
local magnetic field $B$ at
the muon site, and which is proportional to the
positron asymmetry function \cite{steve} $A(t)$.

\section{NVO}

Detailed studies of NVO involving specific heat, 
magnetization, magnetic susceptibility and neutron diffraction
measurements \cite{lawes_mag,kenzelmann} 
reveal a complex magnetic phase diagram which, 
in zero applied magnet field,
is proposed to consist of five phases (See Figs.~\ref{data} and \ref{fit}). 
We describe these below:
\begin{itemize}
\item{Commensurate (C') phase ($2.3 \leq T \leq 4.0$~K): 
An AFM ordered phase where
the staggered magnetism is principally directed
along the $a$-direction. A weak ferromagnetic moment also exists
along $c$.}
\item{Commensurate (C) phase ($ T \leq 2.3$~K): 
The difference between phases C' and C is unclear, 
although we note that they are separated by a 
pronounced peak in the heat capacity\cite{lawes_mag,kenzelmann} 
at $T_{\mathrm{CC'}}=2.3$~K.}
\item{Low temperature incommensurate (LTI) phase 
($4.0 \leq T \leq 6.3$~K): 
The spin structure is elliptically polarized
with both spine and cross-tie spins in the $a$-$b$ plane. 
In this phase ferroelectric order is also
observable, with a spontaneous electrical
polarization parallel to the $b$-direction.}
\item{High temperature incommensurate (HTI) phase 
($6.3 \leq T \leq 9.1$~K): A further
incommensurate phase dominated by a longitudinally modulated
structure with spine spins mainly parallel to the $a$-axis.}
\item{Paramagnetic (PM) phase ($T > 9.1$~K)}
\end{itemize}
The phase transitions at $T_{\mathrm{PH}}=9.1$~K and $T_{\mathrm{HL}}=6.3$~K
are continuous, while that at $T_{\mathrm{LC}}=4.0$~K is 
discontinuous\cite{lawes_mag,kenzelmann}.

Example $\mu^{+}$SR spectra for NVO, measured in 
each of the magnetic phases, are shown in Fig.~\ref{data}. 
In all of the phases occurring below $T_{\mathrm{PH}}=9.1$~K
we observe oscillations 
in the asymmetry spectra. 
The oscillations are characteristic of a quasistatic local
magnetic field at the muon site, which causes a coherent precession of 
the spins of 
those muons with a component of their spin 
polarization perpendicular to
this local field; their presence
provides
strong evidence for the existence of LRO
in these phases, in agreement with previous neutron diffraction 
measurements\cite{lawes_mag,kenzelmann}.
The frequency of the oscillations is given by 
$\nu_{i}=\gamma_{\mu} B_{i}/2 \pi$, where
$\gamma_{\mu}$ is the muon gyromagnetic ratio 
($\equiv 2 \pi \times 135.5$~MHz T$^{-1}$) and $B_{i}$ is the local field at
the $i$th muon site. In the presence of a distribution of
local magnetic fields the oscillations are expected to relax. The nature
of the oscillations and their relaxation differ in different regions of the
phase diagram of NVO, allowing an insight into the local magnetic
field distribution and its correlations in each of the phases. We
describe the muon response to each phase in detail below. 

\subsection{C' and C phases}

\begin{figure}
\begin{center}
\epsfig{file=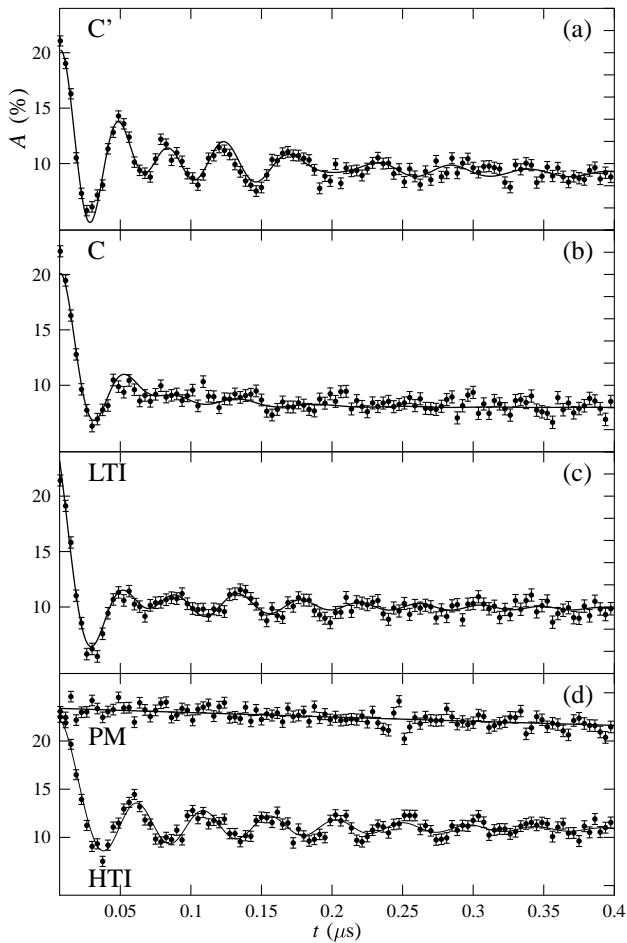,width=8.5cm}
\caption{ZF $\mu^{+}$SR spectra for NVO measured at temperatures
(a) $T=1.60$~K,
(b) 3.85~K,
(c) 4.34~K,
(d) 6.77~K (HTI )and 9.88~K (PM), with fits 
as described in the 
main text.
\label{data}}
\end{center}
\end{figure}

Example spectra measured in the C' and C phases are shown in 
Fig.~\ref{data}(a) and (b).
The measured spectra are found to be 
similar in these two phases and may be described
with the same relaxation function. 
The observed oscillations in these phases demonstrate the existence of
a narrow distribution of the magnitudes of the static local 
magnetic fields at symmetry-related muon sites in the crystal, 
associated with LRO. 
These oscillations are observed at two distinct frequencies
corresponding to two sets of magnetically inequivalent muon sites 
in the material. 
Fitting the oscillations requires the inclusion of constant phase
offsets $\phi_{i}$, which took the values 
$\phi_{1}=-76.2^{\circ}$ and $\phi_{2}=-64.4^{\circ}$ (see
Eq.~\ref{phaseI} below). 

A successful model of the measured spectra also requires the inclusion 
of a purely relaxing
component $\exp(-\lambda_{3} t)$. 
Irrespective of crystal orientation, 
the fact that the muon couples to dipole fields means that, in
addition to the large magnetic field components directed perpendicular
to the muon spin, there may also exist components parallel to the 
muon spin. These will give rise to non-oscillatory components in the
$\mu^{+}$SR spectra.
An exponentially relaxing component will often arise
due to dynamic fluctuations in the local magnetic field distribution
 experienced by the muon ensemble\cite{hayano}. 

The spectra were well described across the measured temperature range
with the resulting functional form
\begin{eqnarray}
\label{phaseI}
A(t) & = & A_{1} \exp(-\lambda_{1} t)  \cos(2 \pi \nu_{1} t +\phi_{1}) \\ \nonumber
& + &  A_{2}  \exp(-\lambda_{2} t)  \cos(2 \pi \nu_{2} t +\phi_{2})  \\ \nonumber
& + & A_{3} \exp(-\lambda_{3} t) +A_{\mathrm{bg}},
\end{eqnarray}
where $A_{\mathrm{bg}}$ represents a constant background contribution from those
muons that stop in the sample holder or cryostat tail. 
This parameter takes the constant value $A_{\mathrm{bg}}=8.46$\% in all
phases of NVO.
The amplitudes were found to be
constant in this temperature regime, taking the values,
$A_{1}=8.52$, $A_{2}=1.99$, $A_{3}=2.57$\%.
The frequencies were found to be in a constant ratio  $\nu_{1}/
\nu_{2}=0.711$ and 
were fixed as such in the fitting procedure. 

\begin{figure*}
\begin{center}
\epsfig{file=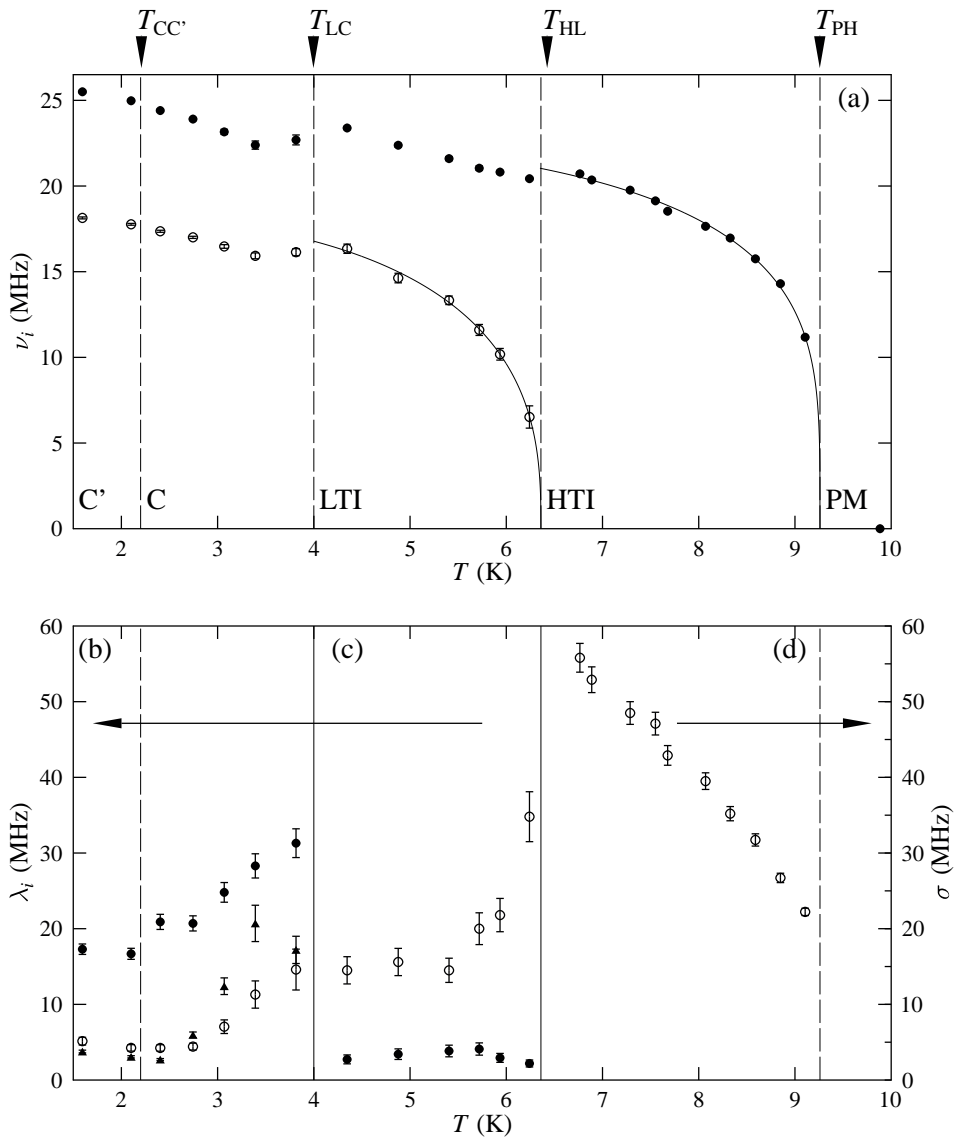,width=13cm}
\caption{Results of fitting the measured data of NVO to 
Eqs.(\ref{phaseI}, \ref{phaseIII}  and \ref{phaseIV}). 
(a) Muon precession frequencies $\nu_{1}$ (closed circles) and
$\nu_{2}$ (open circles). 
(b) and (c) Relaxation rates $\lambda_{1}$ (closed circles), 
$\lambda_{2}$ (open circles) and $\lambda_{3}$ (closed triangles).
(d) Relaxation rate $\sigma$. (Note that solid lines separate the panels;
dashed lines show the temperatures of the reported phase transitions.)
\label{fit}}
\end{center}
\end{figure*}

The results of the fitting procedure are shown in Fig.~\ref{fit}(a) and (b), 
where we see that the frequencies $\nu_{1,2}$ vary continuously across 
$T_{\mathrm{CC'}}$, showing a tendency to decrease with increasing temperature.
We also note the suggestion of a minimum in the frequencies at
around 3.5~K. The relaxation rates $\lambda_{1,2}$ associated with the 
oscillatory components show little 
variation at low temperatures but increase sharply as
$T_{\mathrm{CC'}}$ is approached from
below. 
The relaxation rate 
$\lambda_{3}$ also shows
an increase as $T$ is increased within the C phase but shows a maximum
around $T \sim 3.5$~K.

The need for nonzero phases $\phi_{1,2}$ is probably due to the difficulty 
in modelling
the early time part of the asymmetry spectrum. More significantly, 
the presence of nonzero phases is often a signature of an incommensurate
component of the magnetic order (see below). A small incommensurate
contribution was identified from neutron diffraction measurements
\cite{lawes_mag} on cooling, although this was shown to be
metastable and suppressed by magnetic field cycling. 

No dramatic change of behavior is observed at the phase boundary
of the C' and C phases at temperature $T_{\mathrm{CC'}}=2.1$~K.
The C phase is only distinguished from the C' phase in our measurements 
by the increased influence of fluctuations in the oscillating components 
(reflected by increased relaxation rates $\lambda_{1,2}$)
and by the maximum in the purely relaxing component 
with relaxation rate $\lambda_{3}$. It should be noted, however, 
that the increase in relaxation rates may be ascribed to the
approach to the phase transition at temperature
$T_{\mathrm{LC}}=4.0$~K rather than to an intrinsic difference between
the C' and C phases. 

\subsection{LTI phase}\label{incom}
A sample spectrum measured in the LTI phase is shown in 
Fig.~\ref{data}(c) where oscillations are again apparent. 
The oscillations in the LTI phase are no longer well modelled by 
simple cosinusoidal oscillations; instead, they are best described 
with the use of zeroth order Bessel functions $J_{0}(2 \pi \nu t)$.
This form is expected in cases of incommensurate
magnetic order\cite{amato} where the magnitude of the
magnetic field at the muon site shows a single wavevector 
modulation $B(\mathbf{r}) = B_{0} \sin (\mathbf{k} \cdot \mathbf{r})$. 
In this case the distribution of magnetic fields at the
muon sites is
\begin{equation}
p(B) = \frac{2}{\pi} \frac{1}{\sqrt{1-(B/B_{0})^{2}}} \label{dist}
\end{equation}
yielding a response from the muon ensemble given by
\begin{equation}
A(t) \sim \int p(B) \cos ( \gamma_{\mu} B t) \mathrm{d} B = J_{0}(2 \pi \nu
t),
\end{equation}
where $\nu= \gamma_{\mu} B_{0}$. Although the spin distribution in
the LTI phase has been shown to be helical\cite{lawes_mag}, the
magnitudes of magnetic fields at the muon sites are 
reasonably described by Eq.(\ref{dist}).

As in the C' and C phases, two oscillatory frequencies are observed,
so two Bessel function components were required to model the
data. The data also included a purely relaxing exponential
component to which we assign relaxation rate $\lambda_{3}$. 
The spectra in this phase were fitted to the resulting relaxation
function
\begin{eqnarray}
 \label{phaseIII}
A(t) &=& A_{1} J_{0}(2 \pi \nu_{1} t)\exp(-\lambda_{1} t) \\ \nonumber
&+& A_{2}J_{0}(2 \pi \nu_{2} t)\exp(-\lambda_{2} t)\\ \nonumber
& + & A_{3} \exp (-\lambda_{3} t)+A_{\mathrm{bg}}.
\end{eqnarray}

As in the C and C' phases, the amplitudes were found to take
constant values (in this phase $A_{1}=7.15$, $A_{2}=8.22$ and $A_{3}=1.86$\%). 
The relaxation rates were allowed to vary, except for $\lambda_{3}$
which was found to be relatively small and 
approximately constant in this temperature range and was fixed at 
a value of $\lambda_{3}=1.10$~MHz. 

The frequency $\nu_{1}$ (Fig.~\ref{fit}(a)) 
decreases slowly in this phase. 
In contrast, $\nu_{2}$ shows a more dramatic variation,
tending to zero as $T \rightarrow T_{\mathrm{HL}}$ in the
manner of a continuous phase transition. 
Fitting the frequency $\nu_{2}$ to the phenomenological form
$\nu(T) = \nu(T_{\mathrm{LC}})
(1-(T/T_{\mathrm{HL}})^{\alpha})^{\beta}$, 
with\cite{note} $\alpha=3$, 
gives $\nu(T_{\mathrm{LC}})= 18.6(4)$~MHz, $\beta=0.36(4)$ and yields
an estimate for the transition temperature $T_{\mathrm{HL}}=6.36(5)$~K.
Our estimate for $T_{\mathrm{HL}}$ is in agreement with that of
the previous magnetic measurements\cite{kenzelmann}. 
The temperature evolution of the relaxation rates $\lambda_{1,2}$
is shown in Fig.~\ref{fit}(c). We see that the relaxation rate
$\lambda_{1}$ (associated with the slowly varying frequency $\nu_{1}$) 
shows little variation across the measured temperature regime. Relaxation rate
$\lambda_{2}$, on the other hand (associated with the decreasing
frequency $\nu_{2}$), increases sharply as
$T_{\mathrm{HL}}$ is approached from below. This sharp rise is also
indicative of a phase transition, and may arise due to the onset of
critical fluctuations in the order parameter. 

\subsection{HTI phase}

An example spectrum measured in the HTI phase is shown in
Fig.~\ref{data}(d). The spectra in this regime 
consist of an oscillating component along with a significant
relaxing component. Zeroth order Bessel functions are no longer effective in
describing the oscillatory component of the spectra,
reflecting the change
in magnetic structure that occurs at $T_{\mathrm{HL}}$.
As the temperature approaches $T_{\mathrm{PH}}$ the
relaxation rate becomes sufficiently small to resolve the
lineshape of the relaxing component, which is found to be Gaussian. 
Formally, a broad distribution of local magnetic fields at the muon
site leads to the occurrence of the Kubo-Toyabe (KT) function\cite{hayano}, 
and this is well approximated by a Gaussian function at early times 
with relaxation rate\cite{ktnote} 
$\sigma= \gamma_{\mu} \sqrt{ \langle (B-\langle B \rangle )^{2}\rangle/2}$.
It is
likely therefore that the relaxing component seen in this regime
reflects a broad distribution of magnetic fields at one
set of muon sites. 
The measured spectra were fitted to the relaxation function
\begin{eqnarray}
\label{phaseIV}
A(t) &=& A_{1} \exp(-\lambda_{1})\cos(2 \pi \nu_{1} t + \phi_{1}) \\ \nonumber
& + & A_{2} \exp(-\sigma^{2} t^{2}) +A_{3} \exp(-\lambda_{3}t) +A_{\mathrm{bg}}
\end{eqnarray}
For the fitting procedure, the amplitudes were found to take constant 
values $A_{1}=4.19$, $A_{2}=10.4$ and $A_{3}=2.73$\%, while the
 relaxation rates $\lambda_{i}$ took the values 
$\lambda_{1}=9.04$~MHz, $\lambda_{3}=0.37$~MHz.

The evolution of $\nu_{1}$ and $\sigma$ are shown in Fig.~\ref{fit}(a) and (d) respectively.
The frequency $\nu_{1}$ 
is seen to decrease with increasing temperature tending to
zero as $T_{\mathrm{PH}}$ is approached from below.
Fitting the frequency $\nu_{1}$ to the form
$\nu(T) = \nu(T_{\mathrm{HL}}) (1-(T/T_{\mathrm{PH}})^{\alpha})^{\beta}$, with $\alpha=3$, 
gives $\nu(0)= 23.1(2)$~MHz, $\beta=0.24(1)$ and yields
an estimate for the transition temperature $T_{\mathrm{PH}}=9.26(3)$~K.
The relaxation rate $\sigma$ is seen to decrease with increasing
temperature.

\subsection{PM phase }

In the PM phase we no longer observe oscillations, with the measured spectra being well described by 
a single relaxing component,
\begin{equation}
\label{phaseV}
A(t)=A_{4} \exp(-\lambda_{4} t) +A_{\mathrm{bg}}, 
\end{equation}
This behaviour is typical of magnetic fluctuations of a paramagnetic material.

\subsection{Discussion}
The fact that there are two magnetically distinct Ni$^{2+}$ positions
in NVO is key to understanding the $\mu^{+}$SR results.
The observation of two distinct, temperature dependent signals in each
of the phases below $T_{\mathrm{PH}}$ is consistent with the existence
of two sets of magnetically inequivalent muon sites in the crystal. 
A clue to the origin of the two sets of muon sites is seen by noting that
one difference between the Ni$^{2+}$ positions in the HTI and LTI 
phases\cite{kenzelmann} 
is that $\nu_{2}$ and the moment on the 
cross-tie spins vanish as the system is warmed through the 
transition at $T_{\mathrm{HL}}$. It is likely, therefore, that
the component of the $\mu^{+}$SR signal with amplitude $A_{2}$
arises due to magnetically similar muon sites that are
strongly coupled to the cross-tie Ni$_{\mathrm{c}}$ sites. Since the frequency
$\nu_{1}$ persists up to a temperature $T_{\mathrm{PH}}$,
it is probable that the component of the signal with amplitude $A_{1}$
arises due to muon sites which are strongly coupled to the
spine Ni$_{\mathrm{s}}$ sites.

In the C' and C phases where there is a nonzero 
moment at both sites
we observe oscillations at two distinct frequencies. The larger of
these, $\nu_{1}$, corresponds to
sites near the Ni$_{\mathrm{s}}$ spins 
and the smaller, $\nu_{2}$, to sites near Ni$_{\mathrm{c}}$ spins. 
We again observe two frequencies in the LTI phase, but here the
cross-tie frequency decreases strongly as $T$ is increased. 
The muon results are therefore unique in allowing us to observe the 
continuous decrease of the magnitude of the magnetic moments at the
cross-tie sites followed by the phase transition
associated with these spins that occurs at $T_{\mathrm{HL}}$.
In the HTI phase there is no quasistatic order at the cross-tie sites,
so instead of oscillations we observe a Gaussian signal,
characteristic of a random distribution of static magnetic fields at
these positions. The order at the spine sites persists, however,
leading to a coexistence of oscillations with the Gaussian
relaxation. Comparing the critical parameters estimated from the behaviour
of the two frequencies, we see that the transition associated with the
Ni$_{\mathrm{c}}$ spins, with $\beta=0.36(4)$, appears more three-dimensional 
than that associated with the Ni$_{\mathrm{s}}$ spins which occurs with
$\beta=0.24(1)$. 

The decrease in the Gaussian relaxation rate $\sigma$ also suggests that
the second moment of the local field probed by the muons at the cross-tie 
sites decreases with increasing temperature in the HTI phase. Since it
is unlikely that this distribution is narrowing with increasing
temperature,
it is likely that the magnitude of the local field distribution
is decreasing. This effect has been observed before in $\mu^{+}$SR
studies where two magnetic subsystems have been found to coexist
in a material\cite{tom}. We see therefore that in the
HTI phase the local field at both the spine and cross-tie sites
decreases with increasing temperature, although in this case only
the spine sites are ordered.

For a polycrystalline sample the amplitude of a component
in the $\mu^{+}$SR spectrum should correspond to the number 
of muon sites coupled to the
spin system giving rise to that signal. In this case we would perhaps
expect $A_{1}$ to be double $A_{2}$ in all phases since there are
twice as many spine sites as cross-tie sites in the XVO structure. 
In a single crystal sample (as was used in this study) this will not
necessarily be the case since there will potentially be 
different components of each local field distribution directed perpendicular to the initial muon spin (giving
rise to oscillations) and parallel (giving rise to a 
non-oscillatory amplitude).
Furthermore, a change in magnetic
structure will potentially alter the proportions 
of local magnetic field directed perpendicular and parallel
to the initial muon-spin direction. 
This effect will be dependent on the exact position of the muon site 
 and hence will
give rise to a nontrivial variation of the amplitudes in different
phases. It is therefore difficult to draw conclusions from the 
component amplitudes in the different phases.

\section{CVO}
\begin{figure}
\begin{center}
\epsfig{file=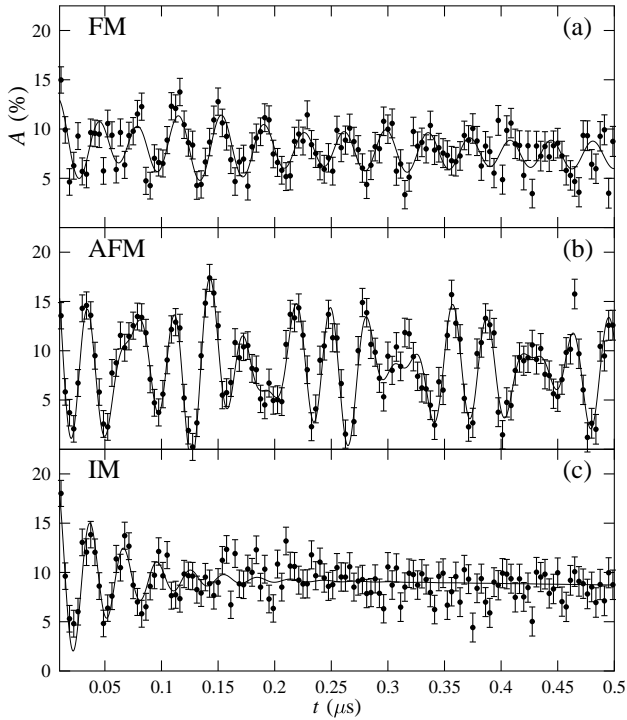,width=8.5cm}
\caption{ZF $\mu^{+}$SR spectra for CVO measured at temperatures
(a) $T=3.50$~K,
(b) 7.30~K,
(c) 9.40~K.
 Fits are shown to Eqs.~(\ref{phaseI}, \ref{phaseIIco} and
 \ref{phaseIIIco}) (see main text).\label{data_co}}
\end{center}
\end{figure}

We now turn our attention to CVO, for which
specific heat\cite{szymczak}, magnetization \cite{szymczak,wilson1} 
and neutron diffraction 
measurements\cite{chen,wilson2} reveal another complex phase diagram,
consisting
of six phases which we describe below (see Figs.~\ref{data_co} 
and \ref{fit_co}). 
\begin{itemize}
\item{Ferromagnetic (FM) phase ($T < 6.2$~K): 
The magnetic structure has 
moments on the spine and cross-tie sites of $2.73~\mu_{\mathrm{B}}$ 
and $1.54~\mu_{\mathrm{B}}$ respectively, ordered along the
$a$-direction\cite{chen}.
}
\item{Antiferromagnetic (AFM') phase  ($6.2 \leq T \leq 6.5$~K): 
A commensurate antiferromagnetic structure exists over a small
temperature range, where the ordering may be
described by a wavevector (0,$\delta$,0), with $\delta=1/3$.}
\item{Incommensurate (IM') phase ($6.5 \leq T \leq 6.8$~K):
The structure is incommensurate in another similarly narrow
temperature range and cannot be described by a simple sinusoidal modulation.}
\item{Commensurate antiferromagnetic phase (AFM) ($6.9 \leq T \leq
    8.6$~K): In this phase the ordering may be described by the
wavevector (0,$\delta$,0), with $\delta=1/2$. The magnetic structure
is characterised by ferromagnetic layers with moments on the spine and
 cross-tie sites of $1.39~\mu_{\mathrm{B}}$ 
and $1.17~\mu_{\mathrm{B}}$ respectively, ordered along the
$a$-direction. These alternate with antiferromagnetic layers where 
the spine site has an ordered moment of $2.55~\mu_{\mathrm{B}}$, while
the cross-tie spins are frustrated and carry no ordered moment.}
\item{Incommensurate  phase (IM) ($8.6 \leq T \leq 11.3$~K): A spin
    density wave state with ordering wavevector (0,$\delta(T)$,0) and 
spin direction along the $a$-direction. The component $\delta(T)$ is
seen to decrease with decreasing temperature from $\delta(11.3$~K$)=0.55$
to $\delta(8.6$~K$)=1/2$.}
\item{Paramagnetic (PM) phase ($T > 11.3$~K).}
\end{itemize}
The transition to the ferromagnetic phase at $T_{\mathrm{A'F}}=6.2$~K 
was found to be discontinuous\cite{chen}.
In our study we have chosen to concentrate on the FM, AFM and IM
phases and example spectra measured in each of these phases
 are shown in Fig.~\ref{data_co}. 
In all of these phases we observe oscillations in the positron
asymmetry spectra. As in the
case of NVO, the nature of the oscillations varies across the phase
diagram, revealing the differences in the local magnetic field
distributions in each phase.

\subsection{FM phase}

\begin{figure}
\begin{center}
\epsfig{file=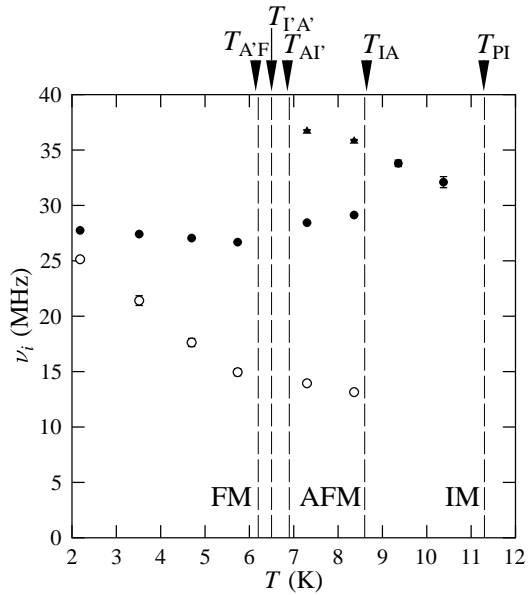,width=8.5cm}
\caption{ Muon precession frequencies $\nu_{1}$ (closed circles),
$\nu_{2}$ (open circles) and $\nu_{3}$ (closed triangles) for CVO,
obtained from fits of
 the measured data to 
Eqs.(\ref{phaseI},\ref{phaseIIco} and \ref{phaseIIIco}).
\label{fit_co}}
\end{center}
\end{figure}

An example spectrum measured in the FM phase is shown in 
Fig.~\ref{data_co}(a).
We find that the $\mu^{+}$SR signal is
comprised of two oscillating components and
a single relaxing component. We may therefor use Eq.(\ref{phaseI})
to fit the measured data in this phase, 
where $A_{\mathrm{bg}}$ takes the value $A_{\mathrm{bg}}=2.0$\% for
CVO. The amplitudes were fixed at the values 
$A_{1}=3.5$, $A_{2}=6.3$ and $A_{3}=1.83$\%,
 while the relaxation rates were $\lambda_{1}=2.0$~MHz,
$\lambda_{2}=0.5$~MHz and $\lambda_{3} \approx 0.1$~MHz.
Nonzero phases were again required to fit the oscillations, with
$\phi_{1}=-63^{\circ}$. More significantly, a successful fit
could only be achieved if $\phi_{2}$ was allowed to 
vary with temperature, following an approximately 
linear variation given by $\phi_{2}= 35.8 T -227$, where $\phi_{2}$ is in
degrees and $T$ in K. 
Fits to Eq.(\ref{phaseI}) yield the frequencies shown in 
Fig.~\ref{fit_co}, where we see that $\nu_{1}$ remains approximately
constant in this phase, while $\nu_{2}$ decreases as $T_{\mathrm{A'F}}$
is approached from below. 

\subsection{AFM phase}

An example spectrum measured in the AFM phase is shown in 
Fig.~\ref{data_co}(b). In this phase the spectra are described
by three frequencies, suggesting three magnetically inequivalent muon
sites. The spectra were fitted to the function
\begin{eqnarray}
\label{phaseIIco}
A(t) &=& A_{1} \exp(-\lambda_{1} t) \cos (2 \pi \nu_{1} t + \phi_{1})
\\ \nonumber
&+& A_{2} \exp(-\lambda_{2} t) \cos (2 \pi \nu_{2} t + \phi_{2})\\ \nonumber
&+& A_{3} \exp(-\lambda_{3} t) \cos (2 \pi \nu_{3} t + \phi_{3}) +A_{\mathrm{bg}}.
\end{eqnarray}
The amplitudes took the constant values
$A_{1}=5.3, A_{2}=4.3$ and $A_{3}=3.0$~\%, while the phases 
took values $\phi_{1}=-34^{\circ}$, $\phi_{2}=-70^{\circ}$ and
$\phi_{3}=-46^{\circ}$. 

The results of fitting the spectra
to Eq.~\ref{phaseIIco} are shown in Fig.~\ref{fit_co}. 
Comparing the frequencies across the phase boundary at
$T_{\mathrm{AF}}$, we see that the frequencies $\nu_{1}$ and
$\nu_{2}$ appear continuous across the phase boundary, while
$\nu_{3}$ emerges at a high frequency.

\subsection{IM phase}
An example spectrum measured in the IM phase is shown in 
Fig.~\ref{data_co}(c). In this phase only one frequency is
observed, along with a purely relaxing component. The data
were fit to the function
\begin{eqnarray}
\label{phaseIIIco}
A(t) &=& A_{1}  \exp(-\lambda_{1} t) \cos (2 \pi \nu_{1} t +
\phi_{1})\\ \nonumber
&+& A_{2} \exp(-\lambda_{2} t) + A_{\mathrm{bg}}.
\end{eqnarray}
In this case the amplitudes were given by $A_{1}=11.8$\%
and $A_{2}=5.6$\%, while $\phi_{1}=-93^{\circ}$ and $\lambda_{2}=0.32$~MHz.

\subsection{Discussion}
The model for NVO of two sets of muon sites, with each strongly coupled
to one of the two inequivalent magnet positions,
also explains the results of our measurements on CVO. 

In the FM phase, where there is an ordered moment on both spine
(2.73~$\mu_{\mathrm{B}}$) and cross tie (1.54~$\mu_{\mathrm{B}}$) 
spins we obtain two oscillations, where we might expect that
the larger frequency $\nu_{1}$ arises from
those muon sites lying close to the Co$_{\mathrm{s}}$ spins
and the smaller $\nu_{2}$ from muon sites near Co$_{\mathrm{c}}$
spins.
The AFM phase in CVO is unusual in that all spine sites are no longer 
equivalent and all cross tie sites are also no longer equivalent. 
Instead we have alternating
ferromagnetically (FM) and AFM coupled staircase layers. Within the
FM coupled layers all Co$_{\mathrm{s}}$ spins (1.39~$\mu_{\mathrm{B}}$)
and Co$_{\mathrm{c}}$ spins (1.17~$\mu_{\mathrm{B}}$) 
are aligned along the $a$-direction\cite{chen}. 
Within the AFM coupled layers, the spines 
 of Co$_{\mathrm{s}}$ spins (2.55~$\mu_{\mathrm{B}}$)
have ordered moments that lie 
parallel or antiparrallel to
$a$ (alternating along the staircase in the $c$-direction), 
while Co$_{\mathrm{c}}$ spins have no ordered moment.
This leads, from the muon's 
point of view, to four inequivalent magnetic environments
(since the muon is insensitive to the direction of the spine
spin ordering). The
disordered Co$_{\mathrm{c}}$ spins within the AFM layers do not lead
to a resolvable relaxation of the muon spin. It is likely that these
fluctuating moments are outside the muon time window and therefore
are motionally narrowed from the spectrum. The three ordered
sites then give rise to the three oscillatory signals observed. From
the size of the moments found from the neutron studies\cite{chen}, 
we assign
$\nu_{1}$ to spine sites within the FM layers, $\nu_{2}$ to
cross-tie sites within the FM layers and the larger frequency
$\nu_{3}$ to the spine sites within the AFM layers. 

The variation of $\phi_{2}$ in the FM phase as the AFM
phase is approached
suggests that there is a continuous 
change in the distribution of local fields at the 
muon sites near Co$_{\mathrm{c}}$ spins as a function
of temperature. This is most likely due to the nature of the transition
between the FM and AFM phases. We see a continuous decrease in the local
magnetic field near the Co$_{\mathrm{c}}$ spins, but it is likely
that this differs in alternating layers depending on whether
the Co$_{\mathrm{c}}$ spin lies on a layer that becomes FM coupled
or AFM coupled in the AFM phases. It follows that Co$_{\mathrm{c}}$
spins on layers becoming AFM coupled will show a far larger decrease in
magnitude than those on FM layers. As a result the two sites become
more inequivalent with increasing temperature, changing the
distribution of local magnetic fields and thus causing $\phi_{2}$
to vary. 

In the IM phase, the magnetic structure is incommensurate and described
by (0, $\delta$, 0), with $\delta > 0.5$ for all spins. 
As a result, only one precession frequency is observed in our measurements. 
Since the two subsets of muon site are unlikely to lie
an equal distance from a Co$_{\mathrm{s}}$ spin or a Co$_{\mathrm{c}}$
spin the contributions from them in an incommensurate structure
will be different. The resulting distribution of magnetic
fields at the muon site will be more complicated than that considered
in section \ref{incom} and so we should not expect a signal that
could be well described by a zero order Bessel function. This does not 
appear to be the case in the LTI phase of NVO, where the order of the 
Ni$_{\mathrm{s}}$ and  Ni$_{\mathrm{c}}$ spins
 gives rise to independent components in the $\mu^{+}$SR spectra.

From our measurements CVO appears to be a less clear case
than NVO and it is possible that the situation may
be more complicated than is considered here. Further work
is required to elucidate this system further.

\section{Conclusions}

We have carried out a detailed study of the kagome staircase 
compounds $X_{3}$V$_{2}$O$_{8}$ ($X=$Ni and Co) using implanted
muons. 
Two sets of muon sites occur in each compound, one set near the spine spins 
and one near the cross-tie spins, allowing us to probe the
two spin environments separately. 
Our results lend additional experimental support to
 the proposed models of magnetic structure for both NVO and CVO
and, in addition, allow us
to follow the temperature evolution of the local field
distribution across the magnetic phase diagrams. 
In the case of NVO the continuous phase transitions are associated with
a continuous decrease in one of the two subsets of spins. The
transition at $T_{\mathrm{LC}}$, where both maintain a nonzero
value, is discontinuous and is
manifested in the spin dynamics of the local magnetic fields. In the
HTI phase, the local field at both the spine and cross-tie sites
decreases with increasing temperature, although in this case only
the spine-site spins are ordered.
For CVO the evolution of the separate subsets of spins is
more complex but provides additional experimental evidence
for  the magnetic transitions
deduced from previous studies.

\acknowledgments 

Part of this work was carried out at the Swiss Muon Source, 
Paul Scherrer Institute, Villigen, Switzerland. We thank Alex Amato
 for technical assistance. This work is supported by the EPSRC. 
T.L.\ acknowledges support from the Royal Commission for the Exhibition
of 1851.


\begin{thebibliography}{99}
\bibitem{moessner}
R. Moessner, Can. J. Phys. {\bf 79} 1283 (2001);
A.P. Ramirez, Annu. Rev. Mater. Sci. {\bf 24} 453 (1994).

\bibitem{sachdev}
S. Sachdev, Phys. Rev. B {\bf 45}, 12377 (1992).

\bibitem{huse}
D.A. Huse and A.D. Rutenberg, Phys. Rev. B {\bf 45}, 7536 (1992).

\bibitem{ramirez}
A.P. Ramirez, G.P. Espinosa and A.S. Cooper, Phys. Rev. Lett. {\bf 64}
2070, (1990).

\bibitem{lee}
S.-H. Lee, C. Broholm, G. Aeppli, T.G. Perring, B. Hessen and
A. Taylor, Phys. Rev. Lett. {\bf 76}, 4424 (1996).

\bibitem{wills}
A. S. Wills, Can. J. Phys. {\bf 79}, 1501 (2001).

\bibitem{rogado}
N. Rogado, G. Lawes, D.A. Huse, A.P. Ramirez and R.J. Cava, Solid
State Commun. {\bf 124}, 229 (2002).

\bibitem{rogado2}
N. Rogado, M.K. Haas, G. Lawes, D.A. Huse, A.P. Ramirez and R.J. Cava, 
J. Phys. Condens. Matter {\bf 15}, 907 (2003).

\bibitem{lawes_mag}
G. Lawes, M. Kenzelmann, N. Rogado, K.H. Kim, G.A. Jorge, R.J. Cava,
A. Aharony, O. Entin-Wohlman, A.B. Harris, T. Yildirim, Q.Z. Huang, 
S. Park, C. Broholm and A.P. Ramirez, 
Phys. Rev. Lett. {\bf 93}, 247201 (2004). 

\bibitem{kenzelmann}
M. Kenzelmann, A.B. Harris, A. Aharony, O. Entin-Wohlman, T. Yildirim,
Q. Huang, S. Park, G. Lawes, C. Broholm, N. Rogado, R.J. Cava,
K.H. Kim, G. Jorge and A.P. Ramirez,
Phys. Rev. B {\bf 74}, 14429 (2006).

\bibitem{szymczak}
R. Szymczak, M. Baran, R. Diduszko, J. Fink-Finowicki, M. Gutowska,
A. Szewczyk and H. Szymczak, Phys. Rev. B {\bf 73}, 94425 (2006).


\bibitem{chen}
Y. Chen, J.W. Lynn, Q. Huang, F.M. Woodward, T. Yildirim, G. Lawes, 
A.P. Ramirez, N. Rogado, R.J. Cava, A. Aharony, O. Entin-Wohlman and
A.B. Harris, Phys. Rev. B, {\bf 74}, 14430 (2006).

\bibitem{wilson1}
N.R. Wilson, O.A. Petrenko and G. Balakrishnan,
arXiv:cond-mat/0610123 

\bibitem{wilson2}
N.R. Wilson, O.A. Petrenko and G. Balakrishnan,
arXiv:cond-mat/0610098


\bibitem{lawes_el}
G. Lawes, A.B. Harris, T. Kimura, N. Rogado,  R.J. Cava, A. Aharony, 
O. Entin-Wohlman, T. Yildrim, M. Kenzelmann, C. Broholm and
A.P. Ramirez, 
Phys. Rev. Lett., {\bf 95}, 87205 (2005).


\bibitem{harris}
A.B. Harris, T. Yildirim, A. Aharony and O. Entin-Wohlman,
Phys. Rev. B, {\bf 73}, 184433 (2006).

\bibitem{khomskii}
D.I. Khomskii, J. Mag. Mag. Mater. {\bf 306} 1 (2006).


\bibitem{musr1}
F. Bert, P. Mendels, A. Olariu, N. Blanchard, G. Collin,
A. Amato, C. Baines and A. D. Hillier, Phys. Rev. Lett.,
{\bf 97} 117203 (2006).

\bibitem{musr2}
P. Dalmas de R\'{e}otier, A. Yaouanc, L. Keller, A. Cervellino,
B. Roessli, C. Baines, A. Forget, C. Vaju, P.C.M. Gubbens,
A. Amato and P.J.C. King,  Phys. Rev. Lett., {\bf 96}, 127202 (2006).

\bibitem{musr3}
X. G. Zheng, H. Kubozono, K. Nishiyama, W. Higemoto, T. Kawae, 
A. Koda and C. N. Xu, Phys. Rev. Lett., {\bf 95}, 057201, (2005). 

\bibitem{musr4}
A. Yaouanc, P. Dalmas de R\'{e}otier, V. Glazkov, C. Marin, P. Bonville, 
J.A. Hodges, P.C.M. Gubbens, S. Sakarya and C. Baines,
Phys. Rev. Lett., {\bf 95}, 047203 (2005).

\bibitem{musr5}
E. Sagi, O. Ofer,  A. Keren and J. S. Gardner,
Phys. Rev. Lett., {\bf 94}, 237202 (2005).

\bibitem{musr6}
D. Bono, P. Mendels, G. Collin, N. Blanchard, F. Bert, A. Amato,
C. Baines, and A.D. Hillier, Phys. Rev. Lett., {\bf 93}, 187201 (2004).

\bibitem{musr7}
A. Fukaya, Y. Fudamoto, I.M. Gat, T. Ito, M.I. Larkin, A.T. Savici,
 Y.J. Uemura, P.P. Kyriakou, G.M. Luke, M.T. Rovers, K.M. Kojima, 
A. Keren, M. Hanawa, and Z. Hiroi, 
Phys. Rev. Lett., {\bf 91}, 207603 (2003).

\bibitem{musr8}
P. Dalmas de R\'{e}otier, A. Yaouanc, P.C.M. Gubbens, 
C.T. Kaiser, C. Baines, and P.J.C. King,
Phys. Rev. Lett., {\bf 91}, 167201 (2003).


\bibitem{balakrishnan}
G. Balakrishnan, O.A. Petrenko, M.R. Lees and D.McK. Paul,
J. Phys. Condens. Matter {\bf 16}, L347 (2004).

\bibitem{steve}
S.J. Blundell, Contemp. Phys. {\bf 40}, 175 (1999).

\bibitem{hayano}
R.S. Hayano, Y. J. Uemura, J. Imazato, N. Nishida, T. Yamazaki and R. Kubo,
 Phys. Rev. B {\bf 20}, 850 (1979).

\bibitem{amato}
A. Amato, Rev. Mod. Phys. {\bf 69},  1119 (1997).

\bibitem{note}
The parameter $\alpha$ models the low temperature behavior
of the frequency evolution. 
As this phase does not persist down to $T=0$
 we are unable to determine
this parameter from the measured data. The choice of $\alpha=3$
(usually chosen to describe the influence
of antiferromagnetic magnons at low temperatures) was found to
provide a satisfactory description of the data in both LTI and HTI
phases, although other parameterizations are possible.

\bibitem{ktnote}
 The recovery of 
asymmetry at late times, characteristic of the
KT function, is often lost due to the presence of slow dynamics
in the local field distribution. As a result, the observed signal is
often only the early time part of the KT function, which is well
described by a Gaussian function. 

\bibitem{tom}
T. Lancaster, S. J. Blundell, D. Prabhakaran, P.J. Baker, W. Hayes 
and F. L. Pratt, Phys. Rev. B {\bf 73}, 184436 (2006).

\end{thebibliography}
\end{document}